\documentclass[10pt]{revtex4}    
\usepackage{amssymb,epsf}
\usepackage{latexsym}
\begin{document}
\title{Generalized second law of thermodynamics in warped DGP braneworld}
\author{Ahmad Sheykhi $^{1,2}$\footnote{sheykhi@mail.uk.ac.ir} and Bin Wang $^{3}$\footnote{wangb@fudan.edu.cn}}
\address{$^1$Department of Physics, Shahid Bahonar University, P.O. Box 76175, Kerman, Iran\\
         $^2$Research Institute for Astronomy and Astrophysics of Maragha (RIAAM), Maragha,
         Iran\\
$^3$  Department of Physics, Fudan University, Shanghai 200433,
China}

\begin{abstract}

We investigate the validity of the generalized second law of
thermodynamics on the $(n-1)$-dimensional brane embedded in the
$(n+1)$-dimensional bulk. We examine the evolution of the apparent
horizon entropy extracted through relation between gravitational
equation and the first law of thermodynamics together with the
matter field entropy inside the apparent horizon. We find that the
apparent horizon entropy extracted through connection between
gravity and the first law of thermodynamics satisfies the
generalized second law of thermodynamics. This result holds
regardless of whether there is the intrinsic curvature term on the
brane or a cosmological constant in the bulk. The observed
satisfaction of the generalized second law provides further
support on the thermodynamical interpretation of gravity based on
the profound connection between gravity and thermodynamics.
\end{abstract}
 \maketitle
 \section{Introduction}\label{Intr}
Inspired by the profound connection between the black hole physics
and thermodynamics, there has been some deep thinking on the
relation between gravity and thermodynamics in general for a long
time. The pioneer work was done by Jacobson who showed that the
gravitational Einstein equation can be derived from the relation
between the horizon area and entropy, together with the Clausius
relation $\delta Q=T\delta S$ \cite{Jac}. Further studies on the
connection between gravity and thermodynamics has been
investigated in various gravity theories \cite{Elin,Cai1,Cai11,Pad1,Pad2,Pad3,Pad4,Pad5,Pad6}. In
the cosmological context, attempts to disclose the connection
between Einstein gravity and thermodynamics were carried out in
\cite{Cai2,Cai3,CaiKim,Fro1,Fro2,Fro3,Fro4,Fro5,verlinde1,verlinde2,verlinde3,verlinde4}. It was shown that the
differential form of the Friedmann equation in the FRW universe
can be written in the form of the first law of thermodynamics on
the apparent horizon. The profound connection provides a
thermodynamical interpretation of gravity which makes it
interesting to explore the cosmological properties through
thermodynamics.

Investigations on the deep connection between gravity and
thermodynamics has recently been extended to braneworld scenarios
\cite{Cai4,Shey1,Shey2}. The motivating idea for disclosing the
connection between the thermodynamics and gravity in braneworld is
to get deeper understanding on the entropy of the black hole in
braneworld. In the braneworld scenarios, gravity on the brane does
not obey Einstein theory, thus the usual area formula for the
black hole entropy does not hold on the brane. The exact analytic
black hole solutions on the brane have not been found so far, so
that the relation between the braneworld black hole horizon
entropy and its geometry is not known. We expect that the
connection between gravity and thermodynamics in the braneworld
can shed some lights on understanding these problems. There are
two main pictures in the braneworld scenario. In the first picture
which we refer as the Randall-Sundrum II model (RS II), a positive
tension 3-brane embedded in  an 5-dimensional AdS bulk and the
cross over between 4D and 5D gravity is set by the AdS radius
\cite{RS1,RS2,Bin}. In this case, the extra dimension has a finite
size. In another picture which is based on the work of Dvali,
Gabadadze, Porrati (DGP model)\cite{DGP,DG}, a 3-brane is embedded
in a spacetime with an infinite-size extra dimension, with the
hope that this picture could shed new light on the standing
problem of the cosmological constant as well as on supersymmetry
breaking \cite{DGP,Wit}. The recovery of the usual gravitational
laws in this picture is obtained by adding to the action of the
brane an Einstein-Hilbert term computed with the brane intrinsic
curvature. The obtained connection between gravity and
thermodynamics in the braneworld shows that the connection is
general and not just an accident in Einstein gravity. The
correspondence of the gravitational field equation describing the
gravity in the bulk to the first law of thermodynamics on the
boundary, the apparent horizon also sheds the light on holography,
since the Friedmann equation persists the information in the bulk
and the first law of thermodynamics on the apparent horizon
contains the information on the boundary. The holographic
description of braneworld scenarios and the enropy function on the
brane have also been explored in \cite{Padi1,Padi2,Sav,James,wang0}.

Besides showing the universality of the connection between gravity
and thermodynamics by expressing the gravitational field equation
into the first law of thermodynamics on the apparent horizon in
different spacetimes, it is of great interest to examine other
thermodynamical principles if the thermodynamical interpretation
of gravity from this correspondence is a generic feature. This is
especially interesting in the braneworld. In the braneworld, the
entropy was extracted through writing the gravitational equation
into the first law of thermodynamics on the apparent
horizon\cite{Shey1,Shey2}. Whether this derived entropy satisfies
general thermodynamical principles is another interesting way to
examine the correctness of the thermodynamical interpretation of
the gravity and the validity of the connection between gravity and
thermodynamics. In this paper we are going to study the
generalized second law of thermodynamics by investigating the
evolution of the apparent horizon entropy deduced through the
connection between gravity and the first law of thermodynamics
together with the matter fields entropy inside the apparent
horizon.  The generalized second law of thermodynamics is a
universal principle governing the universe. Recently the
generalized second law of thermodynamics in the accelerating
universe enveloped by the apparent horizon has been studied
extensively in \cite{wang1,wang2,Shey3}. For other gravity
theories, the generalized second law has also been studied in
\cite{akbar1,akbar2}. In this work we will explore the generalized second
law of thermodynamics in braneworld scenarios, regardless of
whether there is the intrinsic curvature term on the brane or a
cosmological constant in the bulk. If the thermodynamical
interpretation of gravity is correct, the deduced apparent horizon
entropy from the connection between gravitational equation and the
first law of thermodynamics should satisfy the generalized second
law.

Our starting point is the $n$-dimensional homogenous and isotropic
FRW universe on the brane with the metric
\begin{equation}
ds^2={h}_{\mu \nu}dx^{\mu} dx^{\nu}+\tilde{r}^2d\Omega_{n-2}^2,
\end{equation}
where $\tilde{r}=a(t)r$, $x^0=t, x^1=r$, the two dimensional
metric $h_{\mu \nu}$=diag $(-1, a^2/(1-kr^2))$ and $d\Omega_{n-2}$
is the metric of $(n-2)$-dimensional unit sphere. The dynamical
apparent horizon which is the marginally trapped surface with
vanishing expansion, is determined by the relation $h^{\mu
\nu}\partial_{\mu}\tilde {r}\partial_{\nu}\tilde {r}=0$, which
implies that the vector $\nabla \tilde {r}$ is null on the
apparent horizon surface. The apparent horizon was argued as a
causal horizon for a dynamical spacetime and is associated with
gravitational entropy and surface gravity \cite{Hay2,Hay3,Bak}. For the
FRW universe the apparent horizon radius reads
\begin{equation}
\label{radius}
 \tilde{r}_A=\frac{1}{\sqrt{H^2+k/a^2}}.
\end{equation}
The associated surface gravity on the apparent horizon can be
defined as
\begin{equation}
\label{surgra}\label{kappa}
 \kappa =\frac{1}{\sqrt{-h}}\partial_{a}\left(\sqrt{-h}h^{ab}\partial_{ab}\tilde
 {r}\right),
\end{equation}
thus one can easily express the surface gravity on the apparent
horizon
\begin{equation}\label{surgrav}
\kappa=-\frac{1}{\tilde r_A}\left(1-\frac{\dot {\tilde
r}_A}{2H\tilde r_A}\right).
\end{equation}
The associated temperature on the apparent horizon can be
expressed in the form
\begin{equation}\label{Therm}
T_{h} =\frac{|\kappa|}{2\pi}=\frac{1}{2\pi \tilde
r_A}\left(1-\frac{\dot {\tilde r}_A}{2H\tilde r_A}\right).
\end{equation}
where $\frac{\dot{\tilde r}_A}{2H\tilde r_A}<1$ ensures that the
temperature is positive. Recently the Hawking radiation on the
apparent horizon has been observed in \cite{cao} which gives more
solid physical implication of the temperature associated with the
apparent horizon.

\section{GSL of Thermodynamics in RS II braneworld}\label{RS}
Let us start with the Randall-Sundrum (RS II) model in which no
intrinsic curvature term on the brane is included in the action.
The Friedmann equation for $(n-1)$-dimensional brane embedded in
$(n+1)$-dimensional bulk in the RS II model can be written
\cite{Shey1}
\begin{equation}\label{RSFri}
H^2+\frac{k}{a^2}-\frac{2\kappa_{n+1}^2\Lambda_{n+1}}{n(n-1)}-\frac{\mathcal{C}}{a^n}
=\frac{\kappa_{n+1}^4}{4(n-1)^2}\rho^2.
\end{equation}
where \begin{equation}\label{rela}
 \kappa_{n+1}^2=8\pi
 G_{n+1}\, ,\quad
 \Lambda_{n+1}=-\frac{n(n-1)}{2\kappa_{n+1}^2\ell^2},
\end{equation}
$\Lambda_{n+1}$ is the $(n+1)$-dimensional bulk cosmological
constant, $H=\dot{a}/a$ is the Hubble parameter on the brane, and
we assume the matter content on the brane is in the form of the
perfect fluid in homogenous and isotropic universe,
\begin{equation}
 T_{\mu \nu}=(\rho+P)u_{\mu}u_{\nu}+Pg_{\mu \nu},
\end{equation}
where $u^{\mu}$, $\rho$ and $P$ are the perfect fluid velocity
($u^{\mu}u_{\nu}=-1$), energy density and pressure, respectively.
Hereafter we assume that the brane cosmological constant is zero
(if it does not vanish, one can absorb it in the stress-energy
tensor of perfect fluid on the brane). The constant $\mathcal{C}$
comes from the $(n+1)$-dimensional bulk Weyl tensor. Here we are
interested in the flat (Minkowskian) and conformally flat (AdS)
bulk spacetimes, so that the bulk Weyl tensor vanishes and thus we
set $\mathcal{C}=0$ in the following discussions.
\subsection{Brane embedded in Minkowski bulk}\label{minRS}
We begin with the simplest case, namely the Minkowski bulk, in
which $\Lambda_{n+1}=0$. We can rewrite the Friedmann equation
(\ref{RSFri}) in the simple form
\begin{equation}
\label{minRS1}
 H^2+\frac{k}{a^2}=\frac{\kappa_{n+1}^4 }{4 (n-1)^2}\rho^2.
 \end{equation}
In terms of the apparent horizon radius, we can rewrite the
Friedmann equation (\ref{minRS1}) on the brane as
\begin{equation}
\label{minRS2}
 \frac{1}{\tilde {r}_{A}}=\frac{4\pi G_{n+1}}{n-1} \rho,
 \end{equation}
where we have used Eq. (\ref{rela}). Now, differentiating equation
(\ref{minRS2}) with respect to the cosmic time and using the
continuity equation
\begin{equation}
\label{Cont}
 \dot{\rho}+(n-1)H(\rho+P)=0,
\end{equation}
we get
\begin{equation} \label{dotr1}
\dot{\tilde{r}}_{A}=4\pi G_{n+1}H(\rho+P){\tilde{r}_{A}^2}.
\end{equation}
One can see from the above equation that $\dot{\tilde{r}}_{A}>0$
provided that the dominant energy condition, $\rho+P>0$, holds. In
our previous work \cite{Shey1}, we showed that the Friedmann
equation can be written in the form of the first law of
thermodynamics on the apparent horizon of the brane
\begin{equation}
dE=T_h dS_{h}+WdV,
\end{equation}
where $W=(\rho-P)/2$ is the matter work density \cite{Hay2,Hay3},
$E=\rho V$ is the total energy of the matter inside the
$(n-1)$-sphere of radius $\tilde{r}_{A}$ on the brane, where
$V=\Omega_{n-1}\tilde{r}_{A}^{n-1}$ is the volume enveloped by
$(n-1)$- dimensional sphere with the area of apparent horizon
$A=(n-1)\Omega_{n-1}\tilde{r}_{A}^{n-2}$ and
$\Omega_{n-1}=\frac{\pi^{(n-1)/2}}{\Gamma((n+1)/2)}$. Using this
procedure we extracted an expression for the entropy at the
apparent horizon on the brane \cite{Shey1}
\begin{eqnarray}\label{entminRS}
S_{h} =\frac{2\Omega_{n-1}\tilde {r}_{A}^{n-1}}{4G_{n+1}}.
\end{eqnarray}
It is worth noting that the entropy obeys the area formula of
horizon in the bulk (the factor $2$ comes from the $\mathbb{Z}_2$
symmetry in the bulk). This is due to the fact that because of the
absence of the negative cosmological constant in the bulk, no
localization of gravity happens on the brane. As a result, the
gravity on the brane is still $(n+1)$-dimensional. Let us now turn
to find out $T_{h} \dot{S_{h}}$:
\begin{equation}\label{TSh}
T_{h} \dot{S_{h}} =\frac{1}{2\pi \tilde r_A}\left(1-\frac{\dot
{\tilde r}_A}{2H\tilde r_A}\right)\frac{d}{dt}
\left(\frac{2\Omega_{n-1}\tilde {r}_{A}^{n-1}}{4G_{n+1}}\right).
\end{equation}
After some simplification and using Eq. (\ref{dotr1}) we get
\begin{equation}\label{TSh1}
T_{h} \dot{S_{h}} =(n-1)\Omega_{n-1} H(\rho+P){\tilde
r_A}^{n-1}\left(1-\frac{\dot {\tilde r}_A}{2H\tilde r_A}\right).
\end{equation}
As we argued above the term $\left(1-\frac{\dot {\tilde
r}_A}{2H\tilde r_A}\right)$ is positive to ensure $T_{h}>0$,
however, in the accelerating universe the dominant energy
condition may violate, $\rho+P<0$. This indicates that the second
law of thermodynamics ,$\dot{S_{h}}\geq0$, does not hold. Then the
question arises, ``will the generalized second law of
thermodynamics, $\dot{S_{h}}+\dot{S_{m}}\geq0$, can be satisfied
on the brane?'' The entropy of matter fields inside the apparent
horizon, $S_{m}$, can be related to its energy $E=\rho V$ and
pressure $P$ in the horizon by the Gibbs equation \cite{Pavon2}
\begin{equation}\label{Gib1}
T_m dS_{m}=d(\rho V)+PdV=V d\rho+(\rho+P)dV,
\end{equation}
where $T_{m}$ is the temperature of the energy inside the horizon.
We limit ourselves to the assumption that the thermal system
bounded by the apparent horizon remains in equilibrium so that the
temperature of the system must be uniform and the same as the
temperature of its boundary. This requires that the temperature
$T_m$ of the energy inside the apparent horizon should be in
equilibrium with the temperature $T_h$ associated with the
apparent horizon, so we have $T_m = T_h$\cite{Pavon2}. This
expression holds in the local equilibrium hypothesis. If the
temperature of the fluid differs much from that of the horizon,
there will be spontaneous heat flow between the horizon and the
fluid and the local equilibrium hypothesis will no longer hold.
Therefore from the Gibbs equation (\ref{Gib1}) we can obtain
\begin{equation}\label{TSm1}
T_{h} \dot{S_{m}} =(n-1)\Omega_{n-1} {\tilde r_A}^{n-2}\dot
{\tilde r}_A(\rho+P)-(n-1)\Omega_{n-1}{\tilde r_A}^{n-1}H(\rho+P).
\end{equation}
To check the generalized second law of thermodynamics, we have to
examine the evolution of the total entropy $S_h + S_m$. Adding
equations (\ref{TSh1}) and (\ref{TSm1}),  we get
\begin{equation}\label{GSL1}
T_{h}(
\dot{S_{h}}+\dot{S_{m}})=\frac{1}{2}(n-1)\Omega_{n-1}{\tilde
r_A}^{n-2}\dot {\tilde r}_A(\rho+P)=\frac{A}{2}(\rho+P) \dot
{\tilde r}_A.
\end{equation}
where $A>0$ is the area of apparent horizon. Substituting $\dot
{\tilde r}_A$ from Eq. (\ref{dotr1}) into (\ref{GSL1}) we get
\begin{equation}\label{GSL11}
T_{h}( \dot{S_{h}}+\dot{S_{m}})=2\pi G_{n+1} A {\tilde r_A}^{2}
H(\rho+P)^2.
\end{equation}
The right hand side of the above equation cannot be negative
throughout the history of the universe. Hence we have $
\dot{S_{h}}+\dot{S_{m}}\geq0$ which guarantees that the
generalized second law of thermodynamics is fulfilled in a region
enclosed by the apparent horizon on the brane embedded in the
Minkowski bulk.

\subsection{Brane embedded in AdS bulk}\label{AdSRS}
In the previous subsection we assumed that the bulk cosmological
constant is absent and hence we saw that no localization of
gravity happens on the brane. Let us now leave that assumption by
taking $\Lambda_{n+1}<0$, which is the case of the real RS II
braneworld scenario. Using Eq. (\ref{rela}) the Friedmann equation
(\ref{RSFri}) can be written as
\begin{equation}\label{RSAd1}
 \sqrt{H^2+\frac{k}{a^2}+\frac{1}{\ell^2}}= \frac{4\pi
 G_{n+1}}{n-1}\rho.
 \end{equation}
In terms of the apparent horizon radius we have
\begin{equation}\label{RSAd2}
 \rho = \frac{n-1}{4\pi
 G_{n+1}}\sqrt{\frac{1}{{\tilde{r}_A}^2}+\frac{1}{\ell^2}}.
\end{equation}
Taking the derivative of the above equation with respect to the
cosmic time  and using the continuity equation (\ref{Cont}), one
gets
\begin{equation}
\label{RSAd3} \label{dotr2} \dot{\tilde{r}}_A=\frac{4\pi
}{\ell}G_{n+1}
H(\rho+P){\tilde{r}_A}^2\sqrt{{\tilde{r}_A}^2+\ell^2}.
\end{equation}
The entropy expression associated with the apparent horizon in the
RS II braneworld with negative bulk cosmological constant can be
obtained as \cite{Cai4,Shey1}
\begin{equation}
\label{entRSAdS1} \label{entRSAdS1}
S_h=\frac{(n-1)\ell
\Omega_{n-1}}{2G_{n+1}}{\displaystyle\int^{\tilde
r_A}_0\frac{\tilde{r}_A^{n-2}
}{\sqrt{\tilde{r}_A^2+\ell^2}}d\tilde{r}_A}.
\end{equation}
After the integration we have
\begin{equation} \label{entRSAdS2}
S_h=\frac{2\Omega_{n-1}{\tilde{r}_A}^{n-1}}{4 G_{n+1}}
 \times
{}_2F_1\left(\frac{n-1}{2},\frac{1}{2},\frac{n+1}{2},
-\frac{{\tilde{r}_A}^2}{\ell^2}\right),
\end{equation}
where ${}_2F_1(a,b,c,z)$ is the hypergeometric function. It is
worth noticing when $\tilde{r}_A \ll\ell$, which physically means
that the size of the extra dimension is very large if compared
with the apparent horizon radius, one recovers the area formula
for the entropy on the brane given in Eq. (\ref{entminRS}). This
is an expected result since in this regime we have a quasi-
Minkowski bulk and we have shown in the previous subsection that
for a RS II brane embedded in the Minkowski bulk, the entropy on
the brane follows the $(n+1)$-dimensional area formula in the
bulk. Next we turn to calculate $T_{h} \dot{S_{h}}$:
\begin{eqnarray}\label{TSh2}
T_{h} \dot{S_{h}} &=&\frac{1}{2\pi \tilde r_A}\left(1-\frac{\dot
{\tilde r}_A}{2H\tilde r_A}\right)\frac{d}{dt}
\left[\frac{\Omega_{n-1}{\tilde{r}_A}^{n-1}}{2G_{n+1}}
 \times
{}_2F_1\left(\frac{n-1}{2},\frac{1}{2},\frac{n+1}{2},
-\frac{{\tilde{r}_A}^2}{\ell^2}\right)\right] \nonumber\\
&=&\frac{1}{2\pi \tilde r_A}\left(1-\frac{\dot {\tilde
r}_A}{2H\tilde
r_A}\right)\frac{(n-1)\ell\Omega_{n-1}}{2G_{n+1}}\frac{{\tilde{r}_A}^{n-2}\dot
{\tilde r}_A}{\sqrt{\tilde{r}_A^2+\ell^2}}.
\end{eqnarray}
Using Eq. (\ref{dotr2}), after some simplification we obtain again
\begin{equation}\label{TShh3}
T_{h} \dot{S_{h}} =(n-1)\Omega_{n-1} H(\rho+P){\tilde
r_A}^{n-1}\left(1-\frac{\dot {\tilde r}_A}{2H\tilde r_A}\right).
\end{equation}
Adding equation (\ref{TShh3}) with Gibbs equation (\ref{TSm1}), we
reach
\begin{equation}\label{GSL2}
T_{h}(
\dot{S_{h}}+\dot{S_{m}})=\frac{1}{2}(n-1)\Omega_{n-1}{\tilde
r_A}^{n-2}\dot {\tilde r}_A(\rho+P)=\frac{A}{2}(\rho+P) \dot
{\tilde r}_A.
\end{equation}
Substituting $\dot {\tilde r}_A$ from Eq. (\ref{dotr2}) into
(\ref{GSL2}) we get
\begin{equation}\label{GSL22}
T_{h}( \dot{S_{h}}+\dot{S_{m}})=\frac{2\pi}{\ell} G_{n+1} A
{\tilde r_A}^{2}\sqrt{\tilde{r}_A^2+\ell^2}\ H(\rho+P)^2.
\end{equation}
The right hand side of the above equation cannot be negative
throughout the history of the universe, which means that $
\dot{S_{h}}+\dot{S_{m}}\geq0$ always holds. This indicates that
the generalized second law of thermodynamics is fulfilled in the
RS II braneworld embedded in the AdS bulk.

\section{GSL of Thermodynamics in DGP braneworld}\label{DGP}
In the previous section, we have studied the validity of the
generalized second law of thermodynamics in RS II braneworlds
embedded in $(n+1)$-dimensional Minkowski and AdS bulks. In this
section, we would like to extend the discussion to the DGP
braneworld in which the intrinsic curvature term on the brane is
included in the action. The generalized Friedmann equation for the
DGP model is given by \cite{Shey1}
\begin{equation}\label{GFri}
\epsilon
\sqrt{H^2+\frac{k}{a^2}-\frac{2\kappa_{n+1}^2\Lambda_{n+1}}{n(n-1)}-\frac{\mathcal{C}}{a^n}}
=-\frac{\kappa_{n+1}^2}{4\kappa_{n}^2}(n-2)(H^2+\frac{k}{a^2})+\frac{\kappa_{n+1}^2}{2(n-1)}\rho,
\end{equation}
where $\kappa_{n}^2=8\pi
 G_{n}$ and $\epsilon=\pm1$. For later convenience we choose
$\epsilon=1$. Taking the limit $\kappa_{n}\to \infty$, while
keeping $\kappa_{n+1}$ finite, the equation (\ref{GFri}) reduces
to the Friedmann equation (\ref{RSFri}) in the RS II braneworld.
Again, we are interested in studying DGP braneworlds embedded in
the Minkowski and AdS bulks, and we set $\mathcal{C}=0$.

\subsection{Brane embedded in Minkowski bulk}\label{minDGP}
In the Minkowski bulk, $\Lambda_{n+1}=0$, and the Friedmann
equation (\ref{GFri}) reduces to the form
\begin{equation}\label{DGmin1}
\sqrt{H^2+\frac{k}{a^2}}
=-\frac{\kappa_{n+1}^2}{4\kappa_{n}^2}(n-2)(H^2+\frac{k}{a^2})+\frac{\kappa_{n+1}^2}{2(n-1)}\rho.
\end{equation}
In terms of the apparent horizon radius, we can rewrite this
equation in the form
\begin{equation}\label{DGmin2}
 \rho = \frac{(n-1)(n-2)}{2\kappa_{n}^2}\frac{1}{{\tilde{r}_A}^2}+\frac{2(n-1)}{\kappa_{n+1}^2}\frac{1}{{\tilde{r}_A}}
 .\end{equation}
 Now, differentiating equation
(\ref{DGmin2}) with respect to the cosmic time and using the
continuity equation we get
\begin{equation} \label{dotr3}
\dot{\tilde{r}}_{A}=4\pi {\tilde{r}_{A}^2}
H(\rho+P)\left(\frac{n-2}{2G_n \tilde{r}_{A}}+
\frac{1}{G_{n+1}}\right)^{-1}.
\end{equation}
One can see from the above equation that $\dot{\tilde{r}}_{A}>0$
provided that the dominant energy condition, $\rho+P>0$, holds.
However this is not always the case in an accelerating universe.
The entropy expression associated with the apparent horizon in the
pure DGP braneworld can be obtained from the connection between
Friedmann equation and the first law of thermodynamics on the
apparent horizon \cite{Shey1}
\begin{eqnarray}\label{entDGmin1}
S_h
&=&\frac{(n-1)\Omega_{n-1}{\tilde{r}_A}^{n-2}}{4G_n}+\frac{2\Omega_{n-1}{\tilde{r}_A}^{n-1}}
{4G_{n+1}}.
\end{eqnarray}
It is interesting to note that in this case the entropy can be
regarded as a sum of two area formulas; one (the first term)
corresponds to the gravity on the brane and  the other (the second
term) corresponds to the gravity in the bulk. This indeed reflects
the fact that there are two gravity terms in the action of DGP
model. Next we turn to calculate $T_{h} \dot{S_{h}}$:
\begin{eqnarray}\label{TSh3}
T_{h} \dot{S_{h}} &=&\frac{1}{2\pi \tilde r_A}\left(1-\frac{\dot
{\tilde r}_A}{2H\tilde r_A}\right)\frac{d}{dt}
\left(\frac{(n-1)\Omega_{n-1}{\tilde{r}_A}^{n-2}}{4G_n}+\frac{2\Omega_{n-1}{\tilde{r}_A}^{n-1}}
{4G_{n+1}}\right) \nonumber\\
&=&\frac{1}{2\pi \tilde r_A}\left(1-\frac{\dot {\tilde
r}_A}{2H\tilde
r_A}\right)(n-1)\Omega_{n-1}\left(\frac{(n-2){\tilde{r}_A}^{n-3}}{4G_n}+\frac{{\tilde{r}_A}^{n-2}}
{2G_{n+1}}\right)\dot {\tilde r}_A.
\end{eqnarray}
Using Eq. (\ref{dotr3}), we obtain
\begin{equation}\label{TSh33}
T_{h} \dot{S_{h}} =(n-1)\Omega_{n-1} H(\rho+P){\tilde
r_A}^{n-1}\left(1-\frac{\dot {\tilde r}_A}{2H\tilde r_A}\right).
\end{equation}
To check the generalized second law of thermodynamics, we have to
examine the evolution of the total entropy $S_h + S_m$. Combining
equations (\ref{TSh33}) with Gibbs equation (\ref{TSm1}), again we
get
\begin{equation}\label{GSL3}
T_{h}( \dot{S_{h}}+\dot{S_{m}})=\frac{A}{2}(\rho+P) \dot {\tilde
r}_A.
\end{equation}
Substituting $\dot {\tilde r}_A$ from Eq. (\ref{dotr3}) into
(\ref{GSL3}) we get
\begin{equation}\label{GSL33}
T_{h}( \dot{S_{h}}+\dot{S_{m}})=2\pi A {\tilde r_A}^{2}
H(\rho+P)^2 \left(\frac{n-2}{2G_n \tilde{r}_{A}}+
\frac{1}{G_{n+1}}\right)^{-1}.
\end{equation}
The right hand side of the above equation is always positive
throughout the history of the universe. Therefore the generalized
second law of thermodynamics $\dot{S_{h}}+\dot{S_{m}}\geq0$ is
fulfilled in the DGP braneworld embedded in the Minkowski bulk.
\subsection{Brane embedded in AdS bulk}\label{AdSDGP}
For the AdS bulk with $\Lambda_{n+1}<0$, we can write the
Friedmann equation (\ref{GFri}) in the form
\begin{equation}\label{DGAdS1}
\sqrt{H^2+\frac{k}{a^2}+\frac{1}{\ell^2}}
=-\frac{\kappa_{n+1}^2}{4\kappa_{n}^2}(n-2)(H^2+\frac{k}{a^2})+\frac{\kappa_{n+1}^2}{2(n-1)}\rho,
\end{equation}
where we have used Eq. (\ref{rela}). In terms of the apparent
horizon radius, this equation can be rewritten into
\begin{equation}\label{DGAdS2}
 \rho = \frac{(n-1)(n-2)}{2\kappa_{n}^2}\frac{1}{{\tilde{r}_A}^2}+\frac{2(n-1)}{\kappa_{n+1}^2}
 \sqrt{\frac{1}{{\tilde{r}_A}^2}+\frac{1}{\ell^2}}.
\end{equation}
If one takes the derivative of the equation (\ref{DGAdS2}) with
respect to the cosmic time, after using Eqs. (\ref{rela}) and
(\ref{Cont}), one gets
 \begin{equation} \label{dotr4}
\dot{\tilde{r}}_{A}=4\pi {\tilde{r}_{A}^2}
H(\rho+P)\left(\frac{n-2}{2G_n \tilde{r}_{A}}+
\frac{\ell}{G_{n+1}\sqrt{{\tilde{r}_A}^2+\ell^2}}\right)^{-1}.
\end{equation}
When the dominant energy condition holds, $\dot{\tilde{r}}_{A}>0$.
The entropy expression associated with the apparent horizon of the
DGP brane embedded in the AdS bulk can be extracted through
relating the Friedmann equation to the first law of thermodynamics
on the apparent horizon \cite{Shey1}
\begin{eqnarray}\label{entDGAdS1}
 S_{h}&=&
(n-1)\Omega_{n-1}{\displaystyle\int^{\tilde
r_A}_0\left(\frac{(n-2){\tilde{r}_A}^{n-3}}{4G_n}+\frac{\ell}{2G_{n+1}}\frac{{\tilde{r}_A}^{n-2}}{\sqrt{{\tilde{r}_A}^2+\ell^2}}
\right)d\tilde{r}_A}.
\end{eqnarray}
After integration it reads
\begin{equation} \label{entDGAdS2}
S_{h}=\frac{(n-1)\Omega_{n-1}{\tilde{r}_A}^{n-2}}{4G_{n}}+\frac{2\Omega_{n-1}{\tilde{r}_A}^{n-1}}{4G_{n+1}
}
 \times
{}_2F_1\left(\frac{n-1}{2},\frac{1}{2},\frac{n+1}{2},
-\frac{{\tilde{r}_A}^2}{\ell^2}\right).
\end{equation}
Again it is interesting to see that in the warped DGP brane model
embedded in the AdS bulk, the entropy associated with the apparent
horizon on the brane has two parts. The first part follows the
$n$-dimensional area law on the brane and the second part is the
same as the entropy expression obtained in RS II model. The
calculation of $T_{h} \dot{S_{h}}$ yields
\begin{eqnarray}\label{TSh4}
T_{h} \dot{S_{h}} &=&\frac{1}{2\pi \tilde r_A}\left(1-\frac{\dot
{\tilde r}_A}{2H\tilde
r_A}\right)(n-1)\Omega_{n-1}\left[\frac{(n-2){\tilde{r}_A}^{n-3}}{4G_n}+\frac{\ell}
{2G_{n+1}}\frac{{\tilde{r}_A}^{n-2}}{\sqrt{{\tilde{r}_A}^2+\ell^2}}\right]\dot
{\tilde r}_A.
\end{eqnarray}
Substituting $\dot {\tilde r}_A$ from Eq. (\ref{dotr4}), we obtain
\begin{equation}\label{TSh44}
T_{h} \dot{S_{h}} =(n-1)\Omega_{n-1} H(\rho+P){\tilde
r_A}^{n-1}\left(1-\frac{\dot {\tilde r}_A}{2H\tilde r_A}\right).
\end{equation}
Adding equation (\ref{TSh44}) to (\ref{TSm1}), one gets
\begin{equation}\label{GSL4}
T_{h}( \dot{S_{h}}+\dot{S_{m}})=\frac{A}{2}(\rho+P) \dot {\tilde
r}_A.
\end{equation}
Inserting $\dot {\tilde r}_A$ from Eq. (\ref{dotr4}) into
(\ref{GSL4}) we reach
\begin{equation}\label{GSL44}
T_{h}( \dot{S_{h}}+\dot{S_{m}})=2\pi A {\tilde r_A}^{2}
H(\rho+P)^2 \left[\frac{n-2}{2G_n \tilde{r}_{A}}+
\frac{\ell}{G_{n+1}\sqrt{{\tilde{r}_A}^2+\ell^2}}\right]^{-1},
\end{equation}
which cannot be negative throughout the history of the universe
and hence the general second law of thermodynamics,
$\dot{S_{h}}+\dot{S_{m}}\geq0$, is always protected on the DGP
brane embedded in the AdS bulk.

\section{ Summary and discussions}\label{sum}
To conclude, we have investigated the validity of the generalized
second law of thermodynamics for the $(n-1)$-dimensional brane
embedded in the $(n+1)$-dimensional bulk. In the braneworld, the
apparent horizon entropy was extracted through the relation
between the Friedmann equation to the first law of thermodynamics
\cite{Shey1}. We have examined the time evolution of the derived
apparent horizon entropy together with the entropy of matter
fields enclosed inside the apparent horizon on the brane. We have
shown that the extracted apparent horizon entropy through the
connection between Friedmann equation and the first law of
thermodynamics satisfies the generalized second law of
thermodynamics, regardless of whether there is the intrinsic
curvature term on the brane or a cosmological constant in the
bulk. The validity of the generalized second law of thermodynamics
on the brane further supports the thermodynamical interpretation
of gravity and provides more confidence on the profound connection
between gravity and thermodynamics.

Finally, we must mention that as one can see from Eqs. (16), (27), (36) and (45), the variation of the horizon entropy
 takes the same form in different braneworlds. This fact  sheds the light on holography.
The details of the shape of the Hubble  parameter (or the
Friedmann equation) differ in different braneworld models; this is
the bulk effect. However, when they project on the boundary, on
the horizon entropy, these differences are simplified, which just
hid in the $H$, $\dot{\tilde{r}}_{A}$ etc, while the information
on the boundary (the horizon entropy) evolves in the same form.
This is similar to the topological black holes, different topology
will not change the entropy form, $S=A/4$, on the black hole
horizon.

\acknowledgments{We thank the anonymous referee for constructive
comments. This work has been supported financially by Research
Institute for Astronomy and Astrophysics of Maragha, Iran. The
work of B. W. was partially supported by NNSF of China, Shanghai
Science and Technology Commission and Shanghai Education
Commission.}

\end{document}